\documentclass[showpacs,twocolumn,prb]{revtex4-1}

\usepackage{fancyhdr}
\usepackage{graphicx}
\usepackage{amsmath}
\makeatletter
\newcommand{\biggg}{\bBigg@{4}}
\newcommand{\ttt}{\tt}
\makeatother
\usepackage{geometry}
\makeatletter
\def\ps@myPS{%
    \def\@oddfoot{\null\hfill\thepage}
    \def\@evenfoot{\thepage}%
    \def\@evenhead{\null\hfil\slshape\leftmark}%
    \def\@oddhead{{\slshape\rightmark}}}%
\makeatother

\begin{document}
\title{Solving for three-dimensional central potentials using matrix mechanics}
\author{B.A. Jugdutt$^1$ and F. Marsiglio$^{1,2}$}
\affiliation{$^1$ Department of Physics, University of Alberta, Edmonton, Alberta, Canada, T6G~2E1
\\
$^2$Physics Division, School of Science and Technology
University of Camerino, I-62032 Camerino (MC), Italy}

\begin{abstract}
Matrix mechanics is an important component of an undergraduate education in quantum mechanics.
In this paper we present
several examples of the use of matrix mechanics to solve for a number of three dimensional problems
involving central forces. These include examples with which the student is familiar, such as the Coulomb
interaction. In this case we obtain excellent agreement with exact analytical methods. More importantly,
other interesting `non-solvable' examples, such as the Yukawa potential, can be solved as well. Much less
mathematical expertise is required for these methods, while some minimal familiarity with the usage
of numerical diagonalization software is necessary.
\end{abstract}

\date{\today} 
\maketitle


\section{introduction}

An important component of the undergraduate training in quantum mechanics is the solution of the three
dimensional Schr\"odinger equation for a particle that experiences a central 
potential, i.e. one in which the potential is a function only of the distance  
from the origin. Then, the Schr\"odinger equation is separable in spherical coordinates,
and the angular part is determined analytically, in terms of the well known spherical harmonics and the associated
Legendre polynomials.\cite{griffiths05} What then remains is the solution of the radial part of the wave function, and
typically some examples are worked out, like the infinite spherical well, the 3D harmonic oscillator, and of course, the hydrogen atom.

The solution to the radial part of the wave function usually requires somewhat advanced mathematics, and tends
to go in one of two ways, either (i) solution by recognition, or (ii) by power series. The first method amounts to
declaring that the radial differential equation is one that has been studied for more than a century, and is `easily'
recognizable as the `insert famous name' Equation, and therefore has `insert famous name' functions as
solutions. Even worse from a student's point of view, is to declare that the solution is a confluent hypergeometric
function with appropriate arguments, and perhaps leave as an exercise which famous name is associated with
which arguments. The second method is a little more satisfying, in that the student ends up constructing the
function that turns out to have a famous name associated with it, but there are often a number of preliminary
steps required, whereby the asymptotic behaviours are `peeled off', so that what remains is a simple polynomial;
this procedure is straightforward to those that are familiar with it, but to a novice in both quantum mechanics and
in differential equations, the process can be somewhat daunting.

This level of mathematics has its benefits, and is of course a required component of a physicist's toolkit. However, 
at this stage of a student's career it can also serve to dampen their enthusiasm for physics.
As educators it is also important to consider that for students who eventually do not pursue a career in 
physics (or mathematics), extensive knowledge of Laguerre polynomials will probably not help them in their future career. Furthermore, this way of solving
problems will not be too helpful when it comes to examining potentials that are {\em not} tractable by either of
these methods.

At the same time, matrix mechanics is generally taught only in the abstract, with real implementations relegated to
more advanced degrees, and usually in the context of many-body physics. 
We therefore suggest a general purpose numerical method for solving problems involving central forces in three dimensions;
the results obtained are necessarily approximate but very accurate. While this method should not replace the teaching of exact
analytical methods referenced above, it provides a tool for learning the use of matrix mechanics methods, and for understanding
the behaviour of the solutions of various potentials that cannot be solved analytically. It has even been used very recently to provide insight to the differences between the $2s$ and $2p$ electrons in atomic lithium.\cite{stacey12}
This method follows that already developed for one dimensional potentials in Ref. \onlinecite{marsiglio09}. The
virtue of this approach is that it requires only linear algebra and integral calculus, topics normally covered in a
student's first year of university studies. The difficult part is that students need to have access to software tools
at some basic level to carry out the linear algebra, and, in some cases, to perform the 
integrations that are required. Our experience has been that this part is difficult 
for some students; however, it is our belief that some familiarity with
Matlab, or Maple, or Mathematica will have broader application for the average student in the long run 
than a knowledge of non-elementary functions.

We begin with an example which can be first addressed by standard methods, the Coulomb potential, specifically
for the hydrogen atom. In this way students can readily check their answers. The Coulomb potential happens to
be one of the most difficult examples to use, however. Because of its long range it supports an infinite number
of bound states; as will be shown below it is impossible to recover all of these through the present approach, but
a careful study of this problem will help to highlight the limitations and subtleties of this approach. Probably
a few tricks could be adopted to circumvent this difficulty, but this would run counter to our goals, as this method
should be generally applicable to any potential that supports bound states.

We will then examine the finite spherical well, and determine for example, the critical depth required for at least
one bound state to exist. This is also known analytically, and so will provide a benchmark for the present method.

Finally, we will examine solutions for the so-called Yukawa potential, a useful potential both in nuclear physics,
where it was used to model meson exchange between nucleons, and in condensed matter physics, where it is
used to model Coulomb interactions whose range has been shortened due to screening. Solutions for this potential
generally require advanced applications of perturbative or variational methods. We will make comparisons
of our results with these.

We should emphasize that we will not comment on numerical methods in this paper. 
We assume that students have access to software that
can compute desired integrals and diagonalize reasonably large matrices. It is assumed that the necessary training
for this procedure is a prerequisite to a student's first course in quantum mechanics.\cite{prerequisite}

\section{Formulation of the Matrix Mechanics Problem}

For a central potential the solution to the time-independent Schr\"odinger equation is separable,
\begin{equation}
\psi(r,\theta,\phi) \equiv R(r) Y_{\ell}^m(\theta,\phi),
\label{separable}
\end{equation}
where we have already written down the solution to the angular part --- it consists of the spherical harmonics,
which are functions of the standard spherical angles. Following the usual procedure, one can replace the radial function $R(r)$ with $u(r) = rR(r)$, and arrive at the so-called radial equation,
\begin{equation}
-{\hbar^2 \over 2m_0} {d^2u(r) \over dr^2} + V_{\rm eff}(r) u(r) = Eu(r),
\label{uofr}
\end{equation}
which is identical to the one dimensional Schr\"odinger equation for a particle of mass $m_0$, 
except that the {\em effective} potential contains an additional so-called centrifugal term,
\begin{equation}
V_{\rm eff}(r) = V(r) + {\hbar^2 \over 2m_0} {\ell(\ell + 1) \over r^2}.
\label{eff}
\end{equation}
Note also that $\int_0^\infty dr |u(r)|^2 = 1$, and that $r$ ranges from $0$ to $\infty$. In addition, because this 
Schr\"odinger equation is for $u(r) \equiv rR(r)$, and $R(r)$ is well-behaved at the origin, then a boundary condition is
that $u(r=0) = 0$.\cite{shankar80}
Following Ref. \onlinecite{marsiglio09} we embed this potential in an infinite square well\cite{square} extending from 
$r=0$ to $r=a$,
where $a$ is some cutoff radius, whose value will influence the results in a manner to be explained below.
\begin{figure}[h!]
\begin{center}
\includegraphics[height=3.4in,width=2.7in]{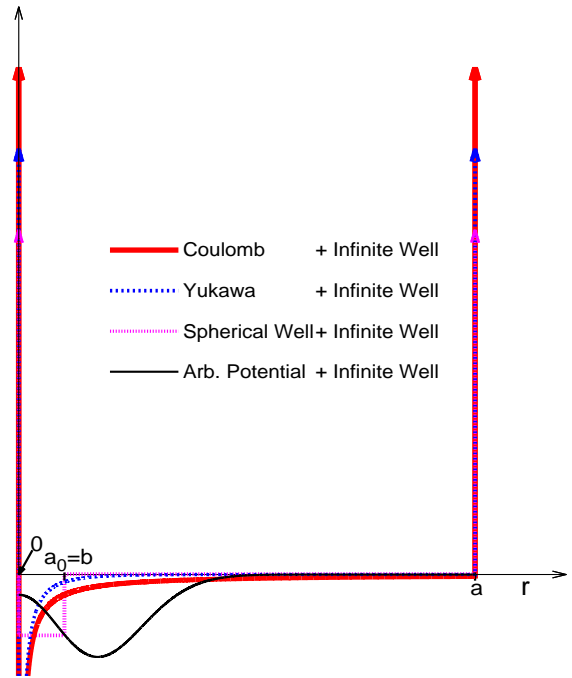} 
\caption{(color online) A plot of the various potentials to be used in this paper, along with the infinite square well
``embedding'' potential placed between $r=0$ and $r=a$. Note that we have used $a_0/a= 1/10$ in the
figure, and $b=a/10$ for the finite spherical well. Also shown is an arbitrarily complicated potential well, to illustrate
the point that this potential poses no further difficulty compared to the others, with this method.}
\label{fig1}
\end{center}
\end{figure}

Figure~1 shows the examples of the potentials we will use in this paper, plotted along with the infinite square well ``embedding''
potential. We also include an arbitrarily complicated potential well shape, to emphasize that this method can
solve for any such potential. 
The rationale for this choice is that the embedding potential allows for a simple set of basis states, which are
simply the eigenstates of the infinite potential well,
\begin{equation}
\phi_n(r) = \sqrt{2 \over a} \sin{\bigl({n \pi r \over a}\bigr)},
\label{basis}
\end{equation}
with eigenvalues
\begin{equation}
E_n^0 = {\pi^2 \hbar^2 n^2 \over 2m_0 a^2}.
\label{basis_energies}
\end{equation}
The embedding potential enforces that the function is zero at the origin, $u(r=0)=0$, but also now requires
the function to vanish at the other wall, $u(r=a)=0$. Such a formulation will work reasonably well for bound
states in attractive potentials, and provides a basis set that is most familiar to students. 
Now we write the radial equation in Dirac notation as
\begin{equation}
[H_0 + V_{\rm eff}]|u\rangle = E|u\rangle,
\label{radial}
\end{equation}
where $H_0$ includes both the kinetic energy and the infinite square well potential, so that
\begin{equation}
H_0 |\phi_n\rangle = E_n^0|\phi_n\rangle.
\label{unpert}
\end{equation}
If we now expand $|u\rangle$ in terms of this basis set,
\begin{equation}
|u\rangle = \sum_{m=1}^\infty c_m |\phi_m\rangle,
\label{expansion}
\end{equation}
and substitute this into Eq. (\ref{radial}), followed by an inner product with each bra $\langle \phi_n|$, we obtain the
matrix equation
\begin{equation}
\sum_{m=1}^\infty H_{nm}c_m = Ec_n,
\label{matrix}
\end{equation}
where the matrix elements are given by
\begin{eqnarray}
&&H_{nm} = \langle \phi_n | (H_0 + V_{\rm eff}) | \phi_m \rangle = \delta_{nm}E_n^0  + \nonumber \\
&&{2 \over a} \int_0^a  dr  \sin{\bigl({n \pi r \over a}\bigr)} \bigl\{V(r) + 
{\hbar^2 \ell(\ell + 1) \over 2m_0 r^2} \bigr\} \sin{\bigl({m \pi r \over a}\bigr)},
\label{matrix_elements}
\end{eqnarray}
and $\delta_{nm}$ is the Kronecker delta function.

Eq. (\ref{matrix_elements}) is readily evaluated for any bound state potential, numerically if need be. We will
begin with the Coulomb potential experienced by an electron of reduced mass $m_0$ near a positively charged
proton,
\begin{equation}
V_{\rm Coul}(r) = - {e^2 \over 4 \pi \epsilon_0}{1 \over r},
\label{vcoulomb}
\end{equation}
where $\mp e$ is the charge of the electron (proton) and $\epsilon_0$ is the vacuum dielectric constant.
Note that both the Coulomb potential and the centrifugal term are singular at the origin, but that the
integrand in Eq. (\ref{matrix_elements}) is not, and varies smoothly, apart from the oscillations of the
sine functions.

Two issues should be raised before we proceed; one is that the matrix size in the eigenvalue problem posed
in Eq. (\ref{matrix}) is infinite. This will be dealt with by utilizing an upper cutoff $n_{max}$, and increasing the
value of this cutoff until the results are converged. For large quantum numbers, the basis states exhibit two
related properties; first, they have increasing energy, and for this reason, may be viewed as more and more irrelevant
for contributions to low energy states. However, concomitantly they have finer spatial resolution. Often it is the
finer spatial resolution that is required to accurately describe a low-lying state, so sometimes a larger basis is
required than one might think if only energy considerations are used. Either way, numerical convergence is attained once 
basis states with very high energy and very sharp spatial resolution are not needed to describe the problem at hand.

The second issue concerns the value of the width of the well, $a$, or equivalently, the cutoff in radial distance.
The natural unit of distance in the Coulomb problem is the Bohr radius, $a_0 \equiv (4 \pi \epsilon_0/ e^2)(
\hbar^2/ m)$, since we know in advance that for the Coulomb problem the bound states decay exponentially
with radial distance $r$. As we shall see, however, exponential decay is not as strong as we would like,
particularly when there is a numerical coefficient in the exponent that stretches out the exponential decay, so that
a cutoff in the radial distance is required to be many times (10-20) the Bohr radius to get very accurate
results for the ground state. For excited states this cutoff will have to be higher, to achieve the same level of accuracy.

For large well widths, however, the basis states deteriorate in spatial resolution. To achieve the same spatial
resolution, therefore, we will need to increase the value of $n_{max}$. We will return to these comments as we
examine particular examples in the following sections.

\section{The Coulomb Potential}

For the Coulomb potential we will focus on $\ell = 0$, to illustrate the method. Note that the integral in 
Eq.~(\ref{matrix_elements}) can be written in terms of the so-called cosine integral. However, in the spirit of
avoiding non-elementary functions (this one is more easily evaluated in the form of the integral
written in Eq.~(\ref{matrix_elements}) anyways), we simply do the integral numerically. Note that in the interest of 
calculating as few of these integrals as is needed ahead of time, it is best to use a trigonometric 
identity\cite{trig} before evaluating the integral. We obtain
\begin{eqnarray}
\langle V_{\rm Coul} \rangle & = & -{e^2 \over 4 \pi \epsilon_0} {2 \over a} \int_0^a dr 
\sin{\bigl({n \pi r \over a}\bigr)} {1 \over r} \sin{\bigl({m \pi r \over a}\bigr)} \nonumber \\
&=& -2{a_0 \over a}E_0 \biggl\{L_1(n+m) - L_1(n-m) \biggr\},
\label{vcoul}
\end{eqnarray}
with
\begin{equation}
L_1(m) \equiv \int_0^1 dx {1 - \cos{(m\pi x)} \over x},
\label{lofm}
\end{equation}
where we used $x \equiv r/a$, added and subtracted unity to the cosines, and, in the second line of 
Eq. (\ref{vcoul}), we adopted the natural energy unit in the problem, $E_0 \equiv \hbar^2/(2m_0a_0^2)$,
which is one Rydberg ($\approx 13.606$ eV). Similar simplification is applicable for the centrifugal term if
needed.

Having decided on a value of $a/a_0$, and a particular maximum size for the matrix, $n_{max}$, it is now
a simple matter of evaluating the matrix elements and substituting into Eq. (\ref{matrix}). We rewrite this
equation in dimensionless units:
\begin{equation}
\sum_{m=1}^{n_{max}} h_{nm}c_m = ec_n,
\label{matrix_dim}
\end{equation}
where $e \equiv E/E_0$, and 
\begin{eqnarray}
h_{nm} &\equiv& {H_{nm} \over E_0} = \delta_{nm} (\pi n a_0/a)^2  \nonumber \\
&-&2{a_0 \over a}\bigl\{L_1(n+m) - L_1(n-m)\bigr\} \nonumber \\
+\ell(\ell + 1) & \bigl({a_0 \over a}\bigr)^2 &\bigl\{L_2(n+m) - 
L_2(n-m)\bigr\},
\label{hmat_dim}
\end{eqnarray}
where
\begin{equation}
L_2(m) \equiv \int_0^1 dx {1 - \cos{(m\pi x)} \over x^2}.
\label{l2ofm}
\end{equation}

Figure~2 shows an example of the calculation with $a/a_0 = 50$ and $n_{max} = 200$. Several characteristics
should be noted. The exact results, $e_{\rm exact} \equiv E_{\rm exact}/E_0 = -1/n^2$, are given by the square symbols. They are of course all negative, though they become more difficult to distinguish from zero as $n$ increases.
\begin{figure}[h!]
\begin{center}
\includegraphics[height=3.4in,width=3.0in,angle=-90]{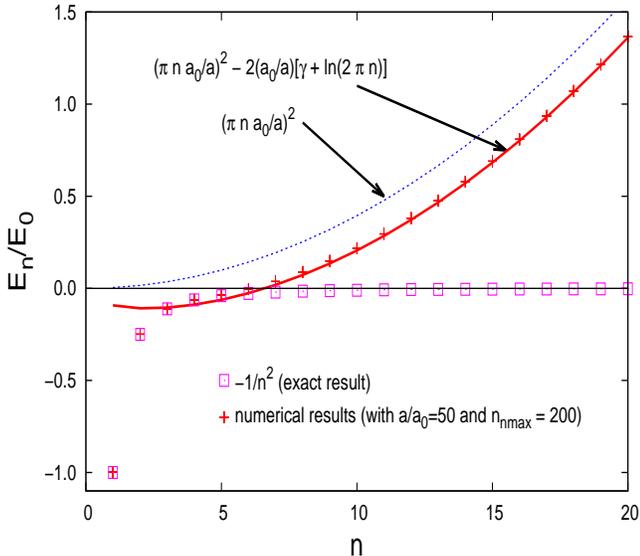} 
\caption{(color online) Energy levels for the Coulomb potential vs. the quantum number $n$ ($\ell = 0$).
Exact analytical results are shown with the squares. Numerical results obtained with an embedding infinite
square well with width $a = 50a_0$, where $a_0$ is the Bohr radius, are shown with cross-hairs. The numerical
results reproduce very accurately the first four bound state energies. Eventually, the energies become positive (unbound) due to the embedding potential, and for very large quantum number $n$, they will
vary as $n^2$ (shown with a blue dotted curve), suitable for an infinite square well of width $a$. A more
accurate result (Eq. (\ref{asymp})) is shown with a red solid curve, and can be derived from perturbation
theory.}
\label{fig2}
\end{center}
\end{figure}
The cross-hairs represent the result of our numerical calculation. Only the first six values of the energy 
are negative, and of these, the first four are very accurate; the remainder become positive and in fact approach the results expected from an infinite square well
potential of width $a$. For such large values of $n$ the Coulomb potential becomes a minor perturbation compared to to the
infinite square well. By examining the diagonal matrix elements only for large values of $n$, one can derive
\begin{equation}
{E_n \over E_0} \approx (\pi n a_0/a)^2 - 2{a_0 \over a}\bigl( \gamma + \ln{(2 \pi n)}\bigr),
\label{asymp}
\end{equation}
where $\gamma \approx 0.5772...$ is Euler's constant. This result is indicated with a curve
in Fig.~2 and provides remarkably accurate results (contrast with the $\sim n^2$ curve also shown), even
for rather low values of $n$. While this analytical result has nothing to do with the pure Coulomb 
potential it does provide an opportunity to illustrate first order perturbation theory.

A summary of the effects of the square well cutoff and the matrix truncation size are best presented
in tabular form, since the differences are so minute. Table~1 shows results as a function of the
matrix size, $n_{max}$, for a given $a/a_0 = 10$. In this instance the third eigenvalue is always positive, i.e.
the square well is narrow enough that what would normally be the third bound state is pushed into the positive
regime by the existence of the outer wall at $r=a$.  Note that the first two bound states, tabulated in Table 1,
do converge to a definite value as $n_{max}$ increases, but that this value is not necessarily the value pertaining to
the Coulomb potential, without the embedding square well potential. In this case, this is especially true
for $E_2/E_0$, which should have a value of $-0.25$, but actually converges to a value of $-0.2256$. 
If we didn't know beforehand that the
expected values were $E_1/E_0 = -1$ and $E_2/E_0 = -1/4$, then the way to check this is to increase the well
width until we achieve convergence in the energies as a function of {\em both} $a/a_0$ and $n_{max}$. Tables
2 and 3 illustrate this process. It is clear that for sufficiently large $a/a_0$ the embedding potential plays no role
(as is desirable) {\em for a sufficiently large matrix size cutoff}. In fact, for a specific cutoff, say $n_{max} = 50$,
then it is clear that as the size of the square well, $a/a_0$, increases, the accuracy for a given energy level {\em
actually decreases}. The reason for this is as stated earlier; for larger $a/a_0$ the same basis state 
(say, $\phi_{50}(r)$) has less spatial resolution than the $50^{th}$ basis state for a smaller value of $a/a_0$.

\begin{table}
\begin{center}
\caption{Table 1. Results for $a/a_0 = 10$.}
\begin{tabular}{|l|l|l|}
\hline
$n_{max}$ & $E_1/E_0$ & $E_2/E_0$\\
\hline
50 & -0.99915 & -0.22546\\
100 & -0.99983 & -0.22558\\
200 & -0.99992 & -0.22560 \\
400 & -0.99993 & -0.22560 \\
\hline 
\end{tabular}
\end{center}
\end{table}
\begin{table}
\begin{center}
\caption{Table 2. Results for $a/a_0 = 20$.}
\begin{tabular}{|l|l|l|l|}
\hline
$n_{max}$ & $E_1/E_0$ & $E_2/E_0$ & $E_3/E_0$ \\
\hline
50 & -0.99428 & -0.24925 & -0.09951\\
100 & -0.99920 & -0.24987 & -0.09979\\
200 & -0.99989 & -0.24996 & -0.09983\\
400 & -0.99999 & -0.24997  & -0.09984\\
\hline 
\end{tabular}
\end{center}
\end{table}

\begin{table}
\begin{center}
\caption{Table 3. Results for $a/a_0 = 50$.}
\begin{tabular}{|l|l|l|l|}
\hline
$n_{max}$ & $E_1/E_0$ & $E_2/E_0$ & $E_3/E_0$ \\
\hline
50 & -0.94114 & -0.24207 & -0.10872\\
100 & -0.98932 & -0.24864 & -0.11071\\
200 & -0.99846 & -0.24981 & -0.11105\\
400 & -0.99980 & -0.24997  & -0.11110\\
800 & -0.99998 & -0.25000  & -0.11111\\
\hline 
\end{tabular}
\end{center}
\end{table}

In summary we have illustrated how matrix mechanics, with a simple square well basis (i.e. a Fourier basis), can
reproduce the bound state energies for the Coulomb potential. By extending the square well width, and appropriately
increasing the matrix size, we converge to the known analytical results. We should also note that most software
packages routinely return the eigenvector along with the eigenvalue. For a given eigenvalue $E_n$, this 
corresponds to
a vector of coefficients $c_m^{(n)}$ for $m=1,2,3...n_{max}$, as in Eq. (\ref{matrix_dim}). With these one can readily compute the corresponding radial wave function, $R_n(r) \equiv u_n(r)/r$, where
\begin{equation}
u_{n}(r) = \sum_{m=1}^{n_{max}} c_m^{(n)} \sqrt{2 \over a} \sin{({m \pi r \over a})}
\label{eigenfn}
\end{equation}
and we have suppressed the index $\ell$ since we have focussed on $\ell = 0$ for the Coulomb potential.
\begin{figure}[h!]
\begin{center}
\includegraphics[height=3.4in,width=3.0in,angle=-90]{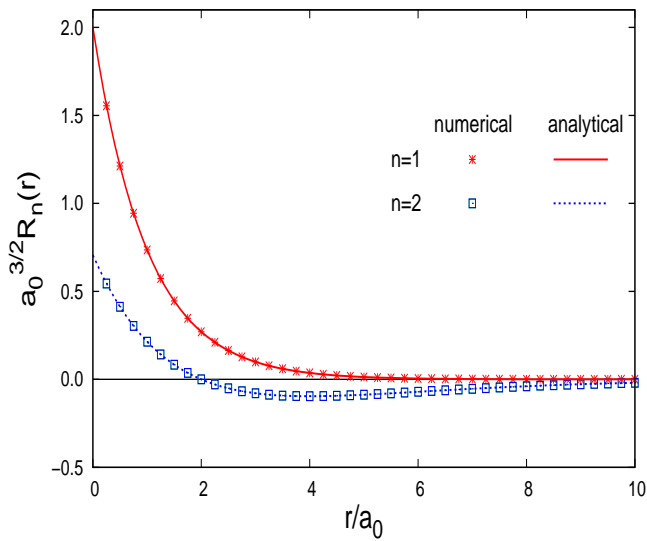} 
\caption{(color online) Numerical results (shown with symbols) for the radial wave 
functions ($n=1$ and $n=2$, both with $\ell = 0$) vs. the radial coordinate, $r$. 
Analytical results are also shown (with curves). Agreement is excellent.}
\label{fig3}
\end{center}
\end{figure}

These are shown in Fig.~3 for the $n=1$ and $n=2$ states, along with the known analytical results,
\begin{eqnarray}
a_0^{3/2} R_{10}(r) &=& 2 \exp{[-r/a_0]} \nonumber \\
a_0^{3/2} R_{20}(r) &=& {1 \over \sqrt{2}} \bigl(1 - {1 \over 2}{r \over a_0} \bigr) \exp{[-r/(2a_0)]};
\label{wavefn}
\end{eqnarray}
the agreement of the numerical results is superb. Note that the scale 
is in units of the Bohr radius, and the decay is complete
over a radial distance of about $10\times$ the Bohr radius. The right hand side of the embedding infinite
square well is at $a=50a_0$, so this is well off-scale on this figure. Because both wave functions have
decayed essentially to zero by this point, the embedding square well disappears from the problem (as is
desired). We have also checked other states, including those with $\ell \neq 0$, and found similar agreement.

\section{Finite Spherical Well}

The finite spherical well (see Fig.~1), with the simple potential,
\begin{equation}
 V_{\text sph}(r) = \begin{cases} - V_0 & \text{if $0 < r < b$,} \\ 
0 & \text{otherwise.}
\end{cases}
\label{spherical_well_potential} 
\end{equation}
has the virtue that, at least for $\ell = 0$, has matrix elements that can be obtained analytically from
Eq. (\ref{matrix_elements}). They are
\begin{equation}
H_{nm} = \delta_{nm}E_n^0 -{V_0 \over \pi} \bigl\{ g(n-m) - g(n+m) \bigr\},
\label{mat_spher}
\end{equation}
where as before $E_n^0 = (\hbar \pi n)^2/[ 2m_0 a^2]$, and the function $g(n)$
is given by
\begin{equation}
g(n) = { \sin{\bigl(n\pi b/a\bigr)} \over n},
\label{gfunc}
\end{equation}
with the $n=0$ case simply given by l'H\^opital's rule.

Students can most readily work with this potential, and make comparisons with known analytical results.
The analytical result in fact requires a graphical solution of a transcendental equation,\cite{griffiths05}
\begin{equation}
\tan{(z + \pi /2)} = \sqrt{\bigl({z_0 \over z}\bigr)^2 - 1},
\label{trans}
\end{equation}
where $z_0 \equiv \pi \sqrt{V_0/E_{1b}^0}$ and $z \equiv \pi \sqrt{(V_0 + E)/ E_{1b}}$. 
Note that we use $E_{1b}^0 \equiv (\hbar \pi)^2/ (2m_0 b^2)$ as the energy scale, as the analytical
solution does {\em not} utilize an embedding infinite potential well of width $a$. Solutions to Eq. (\ref{trans})
are easy to obtain if one simply solves for $z_0$ (and hence $V_0$) in terms of $z$ (and hence $E$). Then,
simple manipulation of Eq. (\ref{trans}) gives the result explicitly, from which a table can be readily constructed,
and then $E$ can be plotted vs. $V_0$.

The spherical potential well can be used to illustrate the important principle that in three dimensions,
a critical depth $V_{0c}$ is required to sustain a bound state, in contrast to the case in one or two 
dimensions. In fact, for a choice $a = 10b$, the embedding potential will have no effect
on the results, {\em except} as the potential depth is varied close to the critical potential. This is because
as this occurs, the spatial extent of the bound state (for $V_0 {{ \atop >} \atop {\sim \atop }} V_{0c}$) increases as 
$V_0 \rightarrow V_{0c}$, until at some point, the embedding potential will `aid' to push the bound state
above zero energy for slightly larger $V_0$ than would actually occur without an embedding infinite square
well.

\begin{figure}[h!]
\begin{center}
\includegraphics[height=3.4in,width=3.0in,angle=-90]{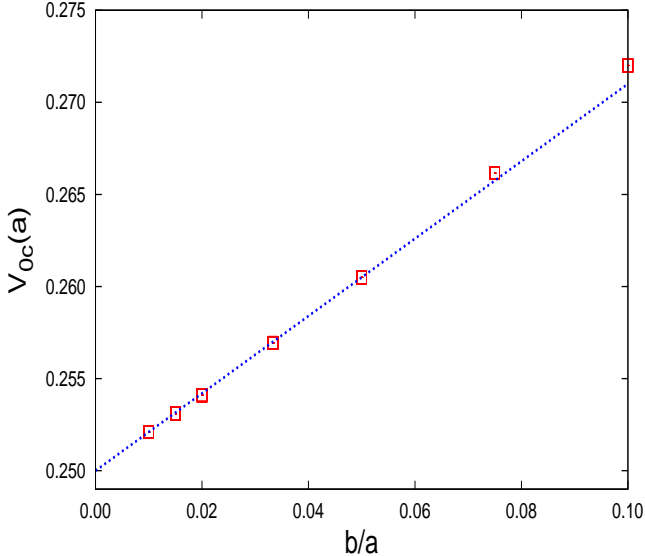} 
\caption{(color online) A plot of $V_{0c}$, the critical value of $V_{0}$, below which a bound state
no longer exists, vs. $b/a$, where $b$ is the radius of the spherical potential, and $a$ is the width
of the embedding infinite square well potential. As $a \rightarrow \infty$ the influence of the embedding
potential is eliminated; the blue line shows that the extrapolated value of $V_{0c} = 1/4$ agrees
with the well known analytical result.}
\label{fig4}
\end{center}
\end{figure}

To discover this with our matrix approach, one would have to increase $a/b$, and determine the value of
the potential for which a negative energy state no longer exists. A plot of these critical values of $V_0$, plotted
versus $b/a$, is shown in Fig.~4. This clearly shows that as $a \rightarrow \infty$, the critical value is
$V_{0c}/E_{1b}^0 = 0.25$, in agreement with what is known analytically. We show this here to address a similar
issue concerning the Yukawa potential in the next section.

\section{The Yukawa Potential}

We now examine an attractive Yukawa potential, which can be written as
\begin{equation}
V_{\rm Yuk}(r) = -A{e^2 \over 4 \pi \epsilon_0}{e^{-\mu r/a_0} \over r},
\label{yuk_potential}
\end{equation}
where $A$ allows us to adjust the strength of the interaction, and $a_0/\mu$ is the screening length, written
in units of the Bohr radius.
Clearly, for $A \rightarrow 1$ and $\mu \rightarrow 0$ we recover the Coulomb interaction discussed in 
Section III. The Yukawa potential has a shorter range than the Coulomb, and therefore has a finite number
of bound states. In what follows we will focus exclusively on the $\ell = 0$ states, and thus have no need
for the centrifugal term, although, as in the case of the unscreened Coulomb interaction, a study of $\ell \ne 0$
states poses no extra difficulty.

Unlike the previous potentials discussed so far, there is no known analytical solution for the Yukawa potential.
Solutions either involve direct numerical solution of the Schr\"odinger differential 
equation,\cite{rogers70} or rely on sophisticated 
numerical procedures centered around the variational 
method\cite{kinderman90,stubbins93,gomes94,luo05} and perturbation theory\cite{vrscay86}; yet another numerical procedure uses 
an expansion technique that uses
special functions connected to Laguerre polynomials.\cite{alhaidari08,bahlouli12} In fact, our 
present methodology is similar in spirit to the so-called J-matrix method\cite{heller74} 
around which these Laguerre polynomial-based methods are developed. 
The virtue of the present approach is in its simplicity; the use of a simple numerical
method based on straightforward matrix diagonalization in a basis which consists of {\em sine} functions makes the work described below readily accessible to undergraduate students.

Following the previous sections, we require the following matrix elements to be used in Eq. (\ref{matrix_dim}):
\begin{eqnarray}
h_{nm} &\equiv& {H_{nm} \over E_0} = \delta_{nm} (\pi n a_0/a)^2  \nonumber \\
&-&2A{a_0 \over a}\bigl\{K_1(n+m) - K_1(n-m)\bigr\} \nonumber \\
+\ell(\ell + 1) & \bigl({a_0 \over a}\bigr)^2 &\bigl\{L_2(n+m) - 
L_2(n-m)\bigr\},
\label{hmat_dim_yuk}
\end{eqnarray}
with
\begin{equation}
K_1(m) \equiv \int_0^1 dx {[1 - \cos{(m\pi x)}] e^{-\mu ax/a_0} \over x},
\label{k1ofm}
\end{equation}
where, as in the case for the Coulomb potential, we have used a unit of energy,
$E_0 \equiv \hbar^2/(2m_0a_0^2)$. Note that the important sections of numerical code
needed to implement this calculation are included in an Appendix; setting $\mu = 0$ allows
one to recover the results for the Coulomb interaction.
\begin{table}
\begin{center}
\caption{Table 4. Ground state energies ($A=1$) in 
units of $E_0$, as compared with those of Ref. (\onlinecite{vrscay86}).}
\begin{tabular}{|l|l|l|}
\hline
$\mu$ & 1s energy  &  Ref. (\onlinecite{vrscay86})\\
\hline
0.10 & -0.814 116 &  -0.814 116...\\
0.20 & -0.653 617 &  -0.653 617...\\
0.40 & -0.396 752  & -0.396 752... \\
0.60 & -0.212 271 &  -0.212 272...\\
0.80 & -0.089 408 & -0.089 409...\\
1.00 & -0.020 552 & -0.020 572...\\
\hline 
\end{tabular}
\end{center}
\end{table}
\begin{figure}[h!]
\begin{center}
\includegraphics[height=3.4in,width=3.0in,angle=-90]{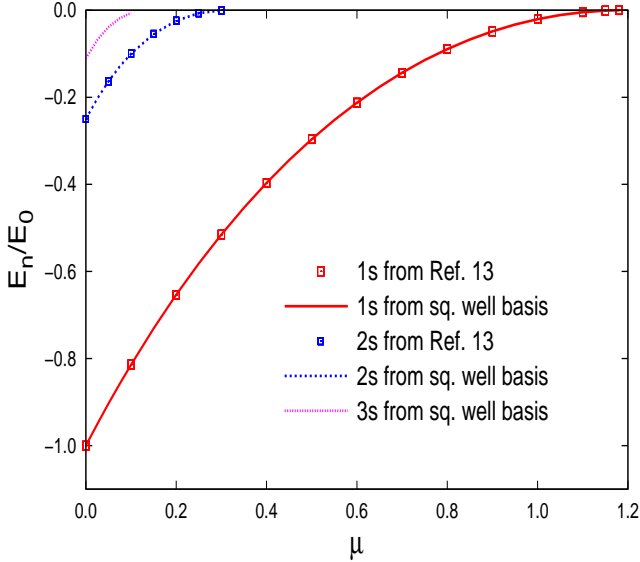}
\caption{(color online) Energy levels for the Yukawa potential (with $A=1$), as a function of the screening parameter, $\mu$. The solid (red), dotted (blue) and dashed (pink) curves show the $1s$, $2s$, and $3s$ levels, respectively. Also shown (with symbols) are results from Ref. (\onlinecite{vrscay86}), with which we
are in excellent agreement. Note the existence of critical values of $\mu_c$, about which we will say
more later.}
\label{fig5}
\end{center}
\end{figure}

The presence of an exponentially decaying factor in the
$K_1(m)$ integral does not make the (numerical) integration any harder than in the Coulomb case;
moreover, for most parameter choices converged results will be obtained without requiring $a/a_0$ to be excessively large. In Fig.~5 we show results for the energies of the
$s$-states; symbols indicate previous results,\cite{vrscay86} which are in excellent
agreement with our own. Table 4 shows more digits for the ground state energies as a
function of the screening parameter $\mu$, and indeed illustrates the remarkable accuracy
of the results of Ref. (\onlinecite{vrscay86}).
We have kept $a/a_0$ and $n_{max}$ fixed at $30$
and $1800$, respectively. This is the reason for the very slight deterioration in our values
for larger $\mu$; we can readily achieve further accuracy by increasing the infinite square well width, but at some point this would become prohibitively time-comsuming. 
As $\mu$ increases (for fixed $A=1$), the bound state energy
approaches zero, and the wave function is more extended, and hence a large value of
$a$ is required to maintain the same accuracy. Note, however, that we are demonstrating that we can achieve any desired accuracy; for example, with $\mu = 0.2$,
we require only $n_{max} \approx 50$ to achieve better than $0.1 \%$ accuracy in the energy. Figure~5 also illustrates graphically how quickly the infinite number of bound states become reduced to a very small number as $\mu$ increases. A critical value, $\mu_c$, beyond which {\em no} bound states exist, clearly exists near $1.2$, about which we will say more below.

\begin{figure}[h!]
\begin{center}
\includegraphics[height=3.4in,width=3.0in,angle=-90]{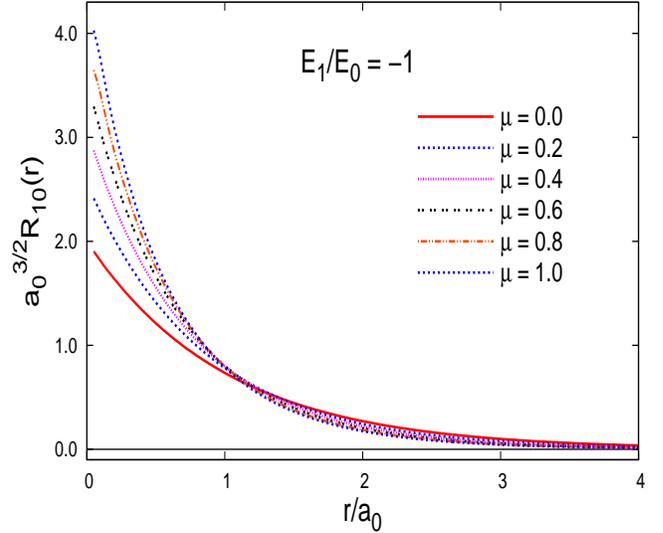} 
\caption{(color online) The ground state wave function vs. radial coordinate, $r$, for various values of
the screening parameter, $\mu$, while $A$ has been increased to keep the ground state energy fixed,
$E_1 = -E_0$. As expected, the wave function becomes increasingly more localized around
the origin, as $\mu$ increases, while keeping the binding energy constant.}
\label{fig6}
\end{center}
\end{figure}

As before one can obtain wave functions;
when screening is present, given the same energy (by increasing $A$ as $\mu$ 
increases) the wave functions are more localized around the origin. Figure~6
shows a comparison of such wave functions; in each case we have adjusted the value
of $A$ to always maintain $E_1(\mu)/E_0 = -1$. The biggest impact occurs near the
origin, as the screened wave functions have considerably more amplitude there. 

\begin{figure}[h!]
\begin{center}
\includegraphics[height=3.4in,width=3.0in,angle=-90]{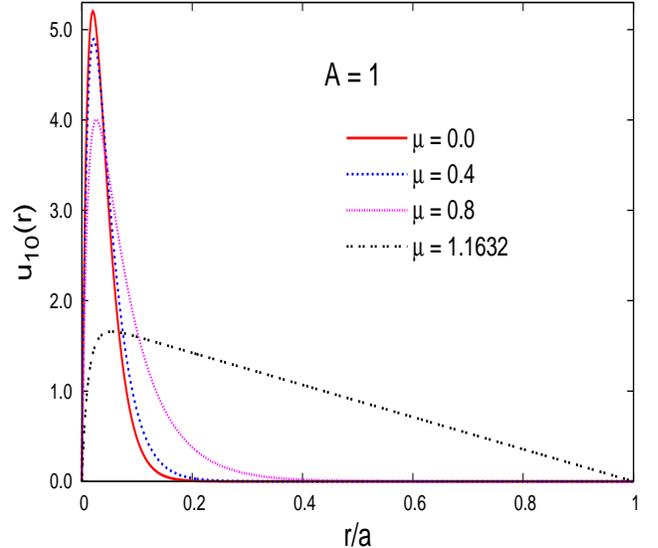} 
\caption{(color online) The weighted ground state wave function, $u(r) \equiv rR_{10}(r)$ 
vs. radial coordinate, $r$, for various values of
the screening parameter, $\mu$, while $A$ is held constant at unity. Now as $\mu$ increases the
binding energy decreases, so the wave function becomes more extended in space. Also shown is
$u(r)$ for $\mu \equiv \mu_c \approx 1.1632$ when $a/a_0 = 50$; now the wave function is
on the verge  of being delocalized over the entire space available, i.e. $0< r < a$.}
\label{fig7}
\end{center}
\end{figure}

Conversely, we can examine how the ground state wave functions evolve 
with increasing $\mu$ with the strength maintained at $A=1$. Then, 
the main effect will be that the energy approaches zero, so that the wave function is increasingly
`less bound' as $\mu$ increases. Thus, even though the interaction is more screened, and therefore
shorter ranged, the wave function will spread out. Figure~7 bears this out; as $\mu$ increases the
wave functions become more extended, as expected. For $\mu \approx \mu_c = 1.1632$ (for the given value of $a/a_0 = 50$ used in Fig.~7) the wave function extends over the entire space allowed by the infinite
square well, and the shape is essentially that of a triangle in $r$:
\begin{equation}
u(r) \approx B\bigl(1 - e^{-\lambda r/a_0}\bigr)\bigl(1 - r/a\bigr),
\label{var_wave}
\end{equation}
where $B$ is determined through normalization. The first factor is required to ensure that $u(r=0)$ is
zero. We find that $\lambda \approx \sqrt{2} \mu_c $ gives a remarkably good fit to the numerically attained wave function (it would be indistinguishable from the numerical result in Fig.~7).

\begin{figure}[h!]
\begin{center}
\includegraphics[height=3.4in,width=3.0in,angle=-90]{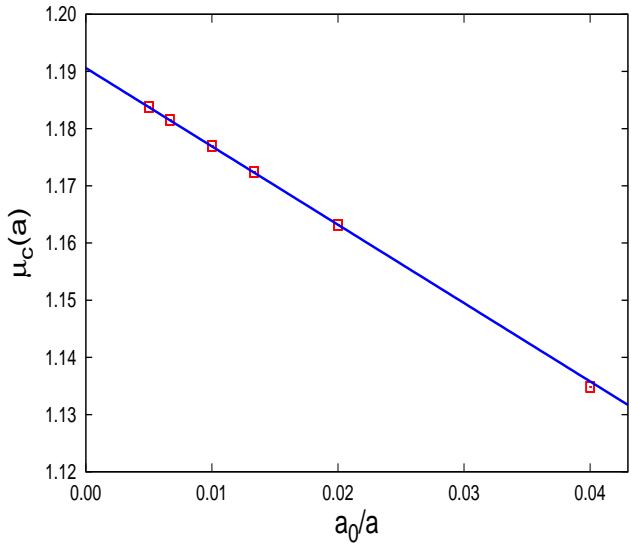} 
\caption{(color online) A plot of $\mu_{c}$, the critical value of $\mu$, above which a bound state
no longer exists in the Yukawa potential, vs. $a_0/a$, where $a_0$ is the Bohr radius, and $a$ is the width of the embedding infinite square well potential. As in the case of the spherical well, as
$a \rightarrow \infty$ the influence of the embedding
potential is eliminated; the blue curve ($\mu_c = 1.1906 - 1.37a_0/a$) shows that the extrapolated value for the critical screening 
parameter is $\mu_{c} \approx 1.1906$, in agreement with previously
established results.}
\label{fig8}
\end{center}
\end{figure}

Finally, we wish to determine how to establish the critical value, $\mu_c$, above which
no bound states exist. We proceed as in the case of the finite spherical well, and
determine the critical value $\mu_c$ as a function of the infinite square well width, $a/a_0$.  The result is plotted in Fig.~8, and we determine, using
the lowest two points (the most reliable) to extrapolate to
$a_0/a \rightarrow 0$, that $\mu_c \approx 1.1906$, in agreement with previous
determinations.\cite{vrscay86,luo05}

\section{Summary}

In this paper we have shown how one can use matrix mechanics, with the simplest of bases, to 
successfully obtain very accurate numerical results for the low-lying levels for essentially any three dimensional potential arising
from central forces that supports bound states. The mathematics required to do this is minimal,
but students must be able to use any number of existing software packages to numerically diagonalize the resulting
matrix. This skillset, non-existent a generation ago, is becoming increasingly useful for an undergraduate
physics degree and beyond.

Both bound state energies and wave functions can be readily obtained with the methodology described
here. By increasing the size of the basis (and, if necessary,
the size of the embedding infinite square well potential to accommodate the spatial spread of the bound
state) one can achieve any desired accuracy. Results were demonstrated for the Coulomb, spherical well, and
Yukawa potentials. We also addressed more difficult issues, such as the existence of critical parameters
(well depth, or screening length) beyond which bound states cease to exist. We were able to reproduce
the textbook result for the critical attractive potential for the spherical well, along with the not so well known
result for the critical screening parameter in the case of the Yukawa potential.

The properties of many other potentials used in the research literature are now accessible to undergraduates;
studies of these potentials are suitable for assignments and/or projects.

\begin{acknowledgments}

This work was supported in part by the Natural Sciences and Engineering Research Council of Canada (NSERC), and by the Teaching and Learning Enhancement Fund (TLEF) and a McCalla Fellowship at the University of Alberta.

\end{acknowledgments}

\appendix
\section{A simple Fortran code to determine the eigenvalues and wave functions for the Yukawa Potential}

Most students will opt to use a high level language like MatLab, Mathematica, or Maple to solve problems as
formulated in this paper. Only a few lines of code are required in this case. We have used $C^{++}$ and Fortran, and here
we write down the key parts of the code required, in Fortran. We use a simple trapezoidal rule to evaluate the
integrals required for the Hamiltonian matrix elements, and call upon two subroutines from Numerical Recipes\cite{press} to diagonalize the matrix.

We use $aa$ to designate the width of the well, while $amu$ is the coefficient $\mu$ in the Yukawa potential, as
written in Eq. (\ref{yuk_potential}). We also use a coefficient $amp$ to vary the strength of the Yukawa potential.
If $\mu = 0$ then we have the Coulomb potential, and $amp$ plays the role of $Z$, the nuclear charge. 
The key points of the code are as follows:

{\ttt 
\noindent c first get and save the needed integrals \\
 \phantom{aa}       gg(0) = 0.0d0 \\
 \phantom{aa}       gg2(0) = 0.0d0 \\
 \phantom{aa}       do 11 n = 1,2*nmax \\
  \phantom{aaaa}        gg(n) = 0.0 \\
  \phantom{aaaa}           gg2(n) = 0.0 \\
    \phantom{aaaa}         do 226 iy = 2,nyy  ! first term is taken care of separately below \\
     \phantom{aaaaaa}          yy = yys + (iy - 1)*dyy \\
    \phantom{aaaaaa}      term = (1.0d0 - dcos(n*pi*yy))/yy \\
    \phantom{aaaaaa}      gg(n) = gg(n) + term*dexp(-amu*aa*yy) \\
    \phantom{aaaaaa}            gg2(n) = gg2(n) + term/yy \\
 226   \phantom{a} continue \\
c contribution from the origin \\
  \phantom{aaaa}         gg(n) = dyy*(gg(n)+ 0.0d0) ! nothing from the origin \\
  \phantom{aaaa}         gg2(n) = dyy*(gg2(n) + 0.5d0*0.5d0*n*n*pi*pi) ! 0.5 from trapezoidal rule \\
 11    \phantom{} continue }

The arrays $gg(n)$ and $gg2(n)$ contain the integrals $K_1(n)$ and $L_2(n)$, respectively, as given
in Eqs. (\ref{k1ofm}) and (\ref{l2ofm}). The next bit of code constructs the matrix elements, $h_{nm}$, as given in
Eq. (\ref{hmat_dim}).

{\ttt 
\noindent c construct the needed matrix elements \\
 \phantom{aa}        do 1 n = 1,nmax \\
 \phantom{aa}          do 2 m = 1,n \\
 \phantom{aaaa}            ag1 = 0.5d0*(gg(m+n) - gg(n-m)) \\
 \phantom{aaaa}            ag2 = 0.5d0*(gg2(m+n) - gg2(n-m)) \\
 \phantom{aaaa}            a(n,m) = 2.0d0*ll2*ag2/(aa*aa) - 4.0d0*amp*ag1*zz/aa \\
 \phantom{aaaa}            if (n.eq.m) a(n,m) = a(n,m) + (pi*n/aa)**2 \\
 \phantom{aaaa}            a(m,n) = a(n,m) \\
2       continue \\
1     continue}

The two dimensional array, $a(n,m)$ now contains the Hamiltonian matrix given by Eq. (\ref{hmat_dim}).
A call to the following two Numerical Recipes\cite{press} routines then diagonalizes and sorts the eigenvalues
and eigenvectors.

{\ttt
\noindent c these two routines, from Numerical Recipes, diagonalize the matrix a(n,m) \\
 \phantom{aa}      call jacobi(a,nmax,dd,vv) \\
 \phantom{aa}      call eigsrt(dd,vv,nmax)}

The eigenvalues are stored in the array $dd(n)$ and the eigenvectors for the $n^{th}$ eigenvalue are stored in
the two dimensional array, $vv(m,n)$. That is, $vv(m,n)$ is the coefficient for the $m^{th}$ basis state to the $n^{th}$
eigenvector. Thus, the next piece of code determines the ground state ($n=1$) and the first excited
state ($n=2$) wave function as a function of $x$.

{\ttt
\noindent c now for the wave function \\
\phantom{aa}      sq2 = dsqrt(2.0d0) \\
\phantom{aa}      do 5 ix = 1,5000 \\
\phantom{aaaa}          xx = ix*0.0002d0 \\
\phantom{aaaa}          bns0 = 0.0d0   ! ground state \\
\phantom{aaaa}          bns1 = 0.0d0   ! first excited state \\
\phantom{aaaa}          do 7 m = 1,nmax \\
\phantom{aaaaaa}             bns0 = bns0 + vv(m,1)*dsin(m*pi*xx) \\
\phantom{aaaaaa}             bns1 = bns1 + vv(m,2)*dsin(m*pi*xx) \\
7       \phantom{aa} continue \\
\phantom{aaaa}          bns0 = bns0*sq2 \\
\phantom{aaaa}          bns1 = bns1*sq2 \\
\phantom{aaaa}          write(71,94)xx,bns0,bns1 \\
5     continue \\
94    format(5x,f9.4,1x,f9.4,1x,f9.4)
}

Similarly, the probabilities can be computed, and the total probability can be checked to sum to unity,
as required in the proper solution to the Schr\"odinger Equation.


\begin{thebibliography}{1}
\bibitem{prerequisite} This is the case at the University of Alberta, and, we believe, at most universities
in the present day.
\bibitem{square} We use the terminology `square well' because we are applying this to the radial equation,
which is a one dimensional equation. Of course, when put in context of the physical problem, this corresponds
to an infinite spherical well.
\end{thebibliography}
\end{document}